\documentclass[twocolumn]{aastex631}
\usepackage{stfloats}
\usepackage{float}
\usepackage{graphicx}
\usepackage{natbib}
\usepackage{hyperref}
\usepackage{amsmath}
\usepackage{autobreak}
\usepackage{multirow}
\usepackage{makecell}
\usepackage{color}
\usepackage{bm}
\usepackage{threeparttable}
\usepackage[figuresright]{rotating}
\usepackage[mathscr]{euscript}
\usepackage[utf8]{inputenc}
\usepackage[T1]{fontenc}
\usepackage{amsmath}
\usepackage{makecell}

\begin{document}

\title{Discovery of a dusty yellow supergiant progenitor for the Type IIb SN~2017gkk}

\correspondingauthor{Ning-Chen Sun}
\email{sunnc@ucas.ac.cn}

\author[0000-0002-3651-0681]{Zexi Niu}
\affiliation{School of Astronomy and Space Science, University of Chinese Academy of Sciences, Beijing 100049, People's Republic of China}
\affiliation{National Astronomical Observatories, Chinese Academy of Sciences, Beijing 100101, China}

\author[0000-0002-4731-9698]{Ning-Chen Sun}
\affiliation{School of Astronomy and Space Science, University of Chinese Academy of Sciences, Beijing 100049, People's Republic of China}
\affiliation{National Astronomical Observatories, Chinese Academy of Sciences, Beijing 100101, China}
\affiliation{Institute for Frontiers in Astronomy and Astrophysics, Beijing Normal University, Beijing, 102206, People's Republic of China}

\author{Jifeng Liu}
\affiliation{National Astronomical Observatories, Chinese Academy of Sciences, Beijing 100101, China}
\affiliation{School of Astronomy and Space Science, University of Chinese Academy of Sciences, Beijing 100049, People's Republic of China}
\affiliation{Institute for Frontiers in Astronomy and Astrophysics, Beijing Normal University, Beijing, 102206, People's Republic of China}
\affiliation{New Cornerstone Science Laboratory, National Astronomical Observatories, Chinese Academy of Sciences, Beijing 100012, People's Republic of China}

\begin{abstract}

Type IIb supernovae are important subclass of stripped-envelope supernovae (SNe), which show H lines only at early times.
Their progenitors are believed to contain a low-mass H envelope before explosion.
This work reports the discovery of a progenitor candidate in pre-explosion Hubble Space Telescope images for the Type IIb SN~2017gkk. 
With detailed analysis of its spectral energy distribution and local environment, we suggest that the progenitor is most likely a yellow supergiant with significant circumstellar extinction and has an initial mass of about 16 $M_\odot$, effective temperature log($T_{\rm eff}/K)=3.72\pm0.08$ and luminosity log($L/L_{\odot})=5.17\pm0.04$. This progenitor is not massive enough to strip envelope through stellar wind, and it
supports an interacting binary progenitor channel and adds to the growing list of direct progenitor detections for Type~IIb SNe.
Future late-time observations will confirm whether this progenitor candidate has disappeared and reveal the putative binary companion that has survived the explosion.

\end{abstract}

\section{Introduction} \label{sec:intro}

Type IIb supernovae (SNe) are characterized by the evolution of H lines in the spectra.
They are a transitional subclass that initially resemble normal Type~II and then evolve to Type~Ib for the disappearance of H lines after the maximum luminosity, usually within 10-20 days \citep{1988Filippenko}.
Some events also show an early shock-cooling peak preceding the main peak powered by $^{56}$Ni decay (especially in the $R$ band), which is distinctive from  light curves for other core-collapse supernovae.
These features indicate progenitors containing a partly stripped H envelope \citep{1993Podsiadlowski}. Theoretically, the single-star progenitor channel requires the fine tuning of initial mass and metallicity \citep{2011Claeys,2017Yoon}. 
Moreover, the fast evolving main peak requires a 
small He core mass (3--4 $M_{\odot}$, \citealp{2012Bersten,2013Benvenuto}) and it is inconsistent with the single progenitor channel which predicts more massive progenitors.
Binary interactions is proposed to be responsible for the low-mass H envelope \citep{1995Nomoto}, and it has been supported by direction detection of progenitors for five Type IIb SNe (SNe~1993J, 2008ax, 2011dh, 2013df, 2016gkg) \citep{1994Aldering,2008Crockett,2011Maund,2014VanDyk,2022Kilpatrick} and their companions \citep{2004Maund,2018Ryder,2019Maund}.
This has also been supported by the nebular-phase spectroscopy \citep{2015Jerkstrand}.
All of them agree with massive stars in interacting binary with $M_{\rm ini}= 13 \sim 17 M_{\odot}$.

In this paper, we report the identification of a progenitor candidate for the Type IIb SN~2017gkk from archival Hubble Space Telescope (HST) images.
We will investigate its progenitor properties with spectral-energy distribution (SED) fitting and age-dating the stellar populations in the vicinity of the SN.
SN~2017gkk exploded in NGC2748 ($z= 0.004923$) and was discovered by \citet{2017Itagaki} on August 31 2017 (MJD 57996) and spectroscopically classified 3 days later \citep{2017Onori}. 
The only documented data for SN~2017gkk is provided by \citet{koivisto2022observational}. They find SN~2017gkk is quite similar with SN~1993J in terms of rising time to the main peak ($\sim$20 days) and $^{56}$Ni mass ($< 0.1 M_{\odot}$), except for redder colors.
The estimated explosion date is August 17 2017 (MJD 57982). 
The velocity of host galaxy is 1507~$\pm$~3~km~s$^{-1}$ with respect to the cosmic microwave background, for a Hubble constant of 73.30~$\pm$~1.04~km~s$^{-1}$~Mpc$^{-1}$ \citep{H0.ref}, this yields a distance of 20.6~Mpc and a distance modulus of 31.5~mag.
\citet{2013Tully,2014Sorce,2016Tully,2020Kourkchi} also estimate the Tully-Fisher distance, ranging from 31.1 to 31.5~mag. We use a weighted mean value of 31.3~mag through the paper.
We adopt a Milky Way extinction of $A_V=0.073$ mag \citep{2011sfd} and a standard extinction law with $R_V = 3.1$ \citep{ebvlaw}.
Datasets and photometry are described in Section \ref{sec:data}. Our analysis is in Section \ref{sec3}. We discuss results and summarize the paper in Section \ref{sec:dis}.

\section{Data and Photometry} \label{sec:data}

The site of SN~2017gkk was imaged by the Wide Field Planetary Camera 2 (WFPC2) in F606W in 1996 and F450W/F814W in 2001, and by the Wide Field Camera 3 (WFC3) in F555W/F814W in 2016, 2019, and 2021 (Table \ref{fig:hst}).  
We downloaded calibrated images from the Mikulski Archive for Space Telescopes\footnote{\url{https://mast.stsci.edu/search/ui/\#/hst}}.
For the WFC3 images, we also re-drizzled them with the \textsc{drrizlepac} package by setting driz$\_$cr$\_$grow = 3 for better cosmic ray removal.
The point-spread-function (PSF) photometry was performed with the \textsc{dolphot} package \citep{2000Dolphin}, where the recommended photometric parameters provided in the user's guide was applied. 
For non-detections, we inserted artificial stars at the stars' positions with different magnitudes; the 3$\sigma$ detection limit was estimated with the magnitude when the detection probability falls to 50\%.
All magnitudes are reported in the Vega system.

\section{Results}\label{sec3}

\subsection{SN Progenitor}\label{sec31}

\begin{figure*} [htbp] 
    \centering %
    \includegraphics[width=\linewidth]{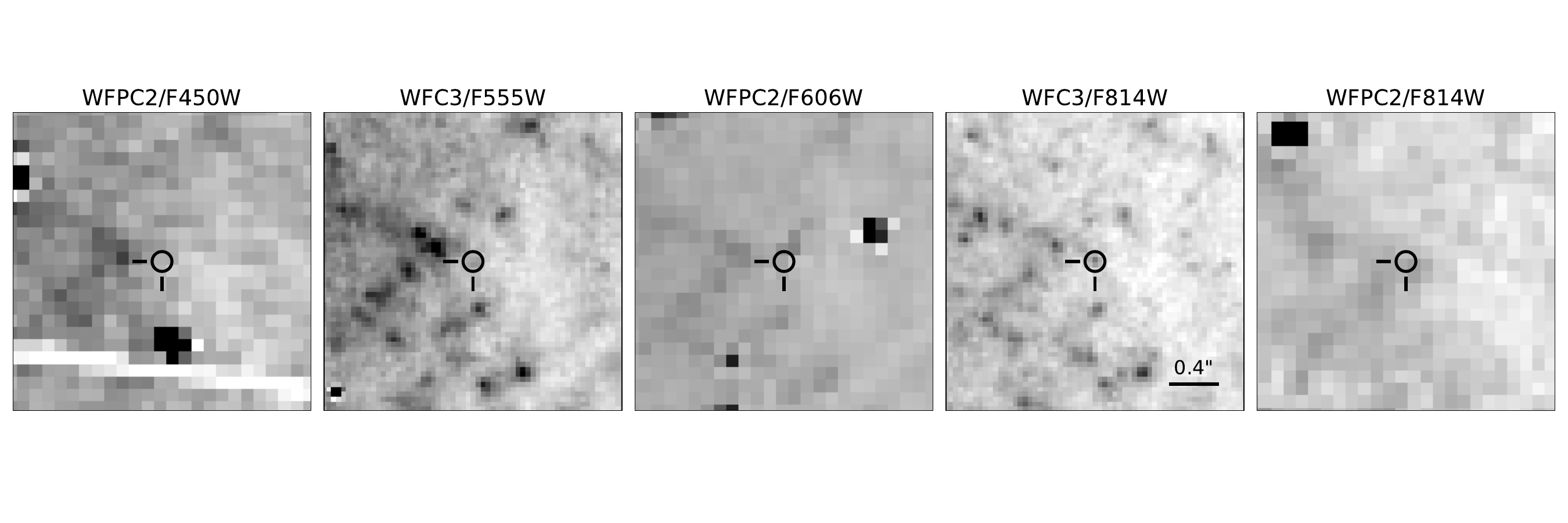}
    \caption{Pre-explosion HST images of the site of SN~2017gkk. All stamps are centered at the SN position and aligned with North up and East to the left. The progenitor candidate is significantly detected in WFC3/F814W and WFC3/F555W images and marginally detected in WFPC2/F606W image.
    It is, however, not detected in WFPC2/F814W and WFPC2/F450W. 
    The bright strip in WFPC2/F450W is because of the bad pixels. 
    \label{fig:hst}}
\end{figure*}

We located the SN site by aligning the pre-explosion images with respect to the image taken in 2019. We used 14 and 7 common objects to align the WFC3 and WFPC2 images, respectively, and reaching differential astrometric uncertainties of 0.318" for 2016 images,  0.412" for 2001 images, and 0.139" for 1996 images. 
Within in the error circle, there is a point source significantly detected in WFC3/F814W to the 9$\sigma$ level and in WFC3/F555W to the 4$\sigma$ level. We also marginally detected a point source in WFPC2/F606W to the 3$\sigma$ level.
No source is found at the SN site in WFPC2/F814W and WFPC2/F450W, and we estimated the detection limits with artificial star tests. 
Figure \ref{fig:hst} plots the pre-explosion images centered at the SN site. 
All datasets and photometry are listed in Table \ref{tab:tab1}.

This photometry are used to determine the effective temperature and  bolometric luminosity of the progenitor candidate by comparing the observed SED with \citet{ck04} stellar atmospheric models. 
Synthetic model magnitudes are calculated with the \textsc{pysynphot} package \citep{pysynphot.ref}, and the Markov Chain Monte Carlo (MCMC) technique \citep{2013emcee} is used to find the best-fitting parameters. 
We note that the extinction of the progenitor remains very uncertain.
The host extinctions for SNe are generally estimated by the equivalent width of Na \textsc{i} D line. However, as shown in \citet{koivisto2022observational}, SN~2017gkk spectra have very weak Na \textsc{i} D line at the redshift of z=0.004923, indicating very little host extinction.
On the other hand, its first spectrum was obtained about 17 days after the explosion, and we can not exclude the possibility that the progenitor was enshrouded by dusty circumstellar material (CSM) before explosion, which was rapidly destroyed by the SN's intense radiation and/or ejecta \citep{2024Li}.  
Therefore, we leave the total (Galactic+host+circumstellar) extinction as a free parameter in the fitting. 

\begin{table}[H]
\centering
\footnotesize
\caption{HST data and photometry at the SN site. \label{tab:tab1}}
    \begin{tabular}{cccccc}
    \hline
    \hline
     Year &  Proposal & Instrument & Filter & Exp.Time & Magnitude \\
     & ID & & & (s) & (mag) \\
\hline
1996 & 	6359 & WFPC2 & F606W  &  600 & 25.73 (0.31) \\
2001 & 9042 &  WFPC2 &  F450W  &  460 & >24.69 \\
 &  &   &  F814W  & 460 & >23.30  \\
2016 & 14668  & WFC3  & F555W  & 710 &  26.58 (0.25)\\
  &    &    & F814W  &  780 & 24.35 (0.11) \\
2019 &  15166  & WFC3  & F555W  & 710 & 23.59 (0.05) \\
  &    &    & F814W  & 780   & 22.87 (0.06)\\
2021 & 16179  & WFC3 & F555W  & 710 & 24.56 (0.07)\\
  &    &    & F814W  & 780 & 23.88 (0.08)  \\
     \hline
\end{tabular}
\begin{tablenotes}
\item[1] PIs: 6359: Stiavelli, Massimo; 9042: Smartt,Stephen J.; 14668, 15166, 16179:  Filippenko, Alex V. 
      \end{tablenotes}
\end{table}

\begin{figure*} 
    \centering 
    \includegraphics[width=\linewidth]{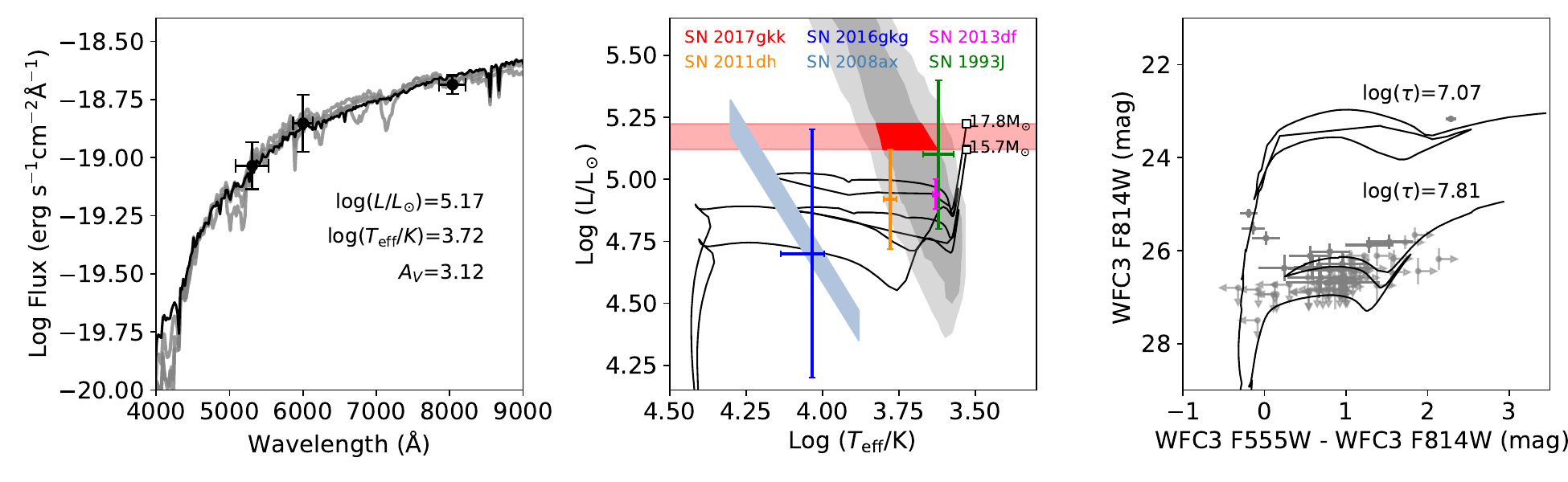}
    \caption{
    \textit{Left}: SED of SN~2017gkk's progenitor candidate. Overplotted are model spectra derived from SED fitting; black solid line with stellar parameters represents result also meeting with environmental analysis (see text for details). 
    \textit{Middle}: Posterior probability contours from the SED fitting for the progenitor candidate in the HRD.
    It can be seen that the effective temperature, luminosity and extinction are highly degenerate.
    The red shade marks the most probable position as inferred from environmental analysis, which derives a likely stellar age of log($\tau$)=7.07$^{+0.03}_{-0.04}$. For comparison, PARSEC single-star stellar isochrones for these ages and the endpoints are shown by black solid lines and open squares marks, respectively. 
    Note that the final stellar luminosity is determined by the final He core mass, which is approximately unaffected by the partial H envelope stripping. 
    Five previously detected Type~IIb SNe progenitors are also indicated (\citealp{2022Kilpatrick,2015Folatelli,2014VanDyk,2011Maund,2004Maund}). 
    \textit{Right}: The CMD for surrounding stars within 150~pc of SN~2017gkk and isochrones of their best-fitting populations.
    \label{fig:progenitor}}
\end{figure*}

The derived extinction, temperature, and luminosity are highly degenerate.
The contour in Figure \ref{fig:progenitor} plots the posterior distributions in the HRD. The lower-extinction models have lower temperatures and luminosities and correspond to lower initial masses, and vice versa.
Note that the final luminosity is tightly correlated with the final core mass \citep{2020Farrell}, and the relation between the final core mass and initial stellar mass is barely affected by the partial loss of H envelope \citep{2021Laplace}; therefore, the initial mass can be derived by comparing the final progenitor luminosity with the single-star isochrones.
Without additional information, the possible initial mass can range from 8 to 19 $M_{\odot}$ depending on the uncertain extinctions.
In the next Section, we shall further analyze stellar populations in the SN vicinity, which allows us to constrain the candidate to be log($T_{\rm eff}/K)=3.72\pm0.08$ and log($L/L_{\odot})=5.17\pm0.04$.

\subsection{SN Environment}\label{sec32}

Most massive stars form in clusters and they evolve so rapidly that they are still close to their birth regions when they explode. The star formation history in the SN environment can provide useful insights into the age of the exploded progenitor.
In order to analyze the resolved stellar populations surrounding the progenitor candidate, we select 68 sources detected on the stacked WFC3 F555W and F814W images with the following criteria\footnote{More detailed explanations of \textsc{dolphot} output catalog can be found in \citealp{2024Weisz}.}:

(1) object type = 1, which selects good star;

(2) signal-to-noise ratio $\ge$ 3;

(3) $-$0.5 $\le$ sharpness $\le$ 0.5, which eliminates sharp objects such as cosmic rays and extended objects such as clusters;

(4) crowding $\le$ 0.5, which selects relatively isolated stars that suffer negligible contamination from neighboring sources;

(5) distance to SN less than 150~pc.

Following the method described in \citet{Maund2016} and \citet{Sun2021}, we performed a hierarchical Bayesian approach based on the \textsc{parsec} v1.2S stellar isochrones \citep{parsec.ref} with solar metallicity to analyze the stellar populations.
Each detected star is assumed to be single stars or non-interacting binaries and they constitute stellar populations with different mean log-ages.
We did not consider interacting binaries, which could be very complicated and is beyond the scope of this work. 
Each model population follows a \citet{imf.ref} initial mass function, and has a 50\% binary fraction with a flat distribution of primary-to-secondary mass ratio.
We note that the binary fraction and the mass ratio distribution are very inconclusive and have complex relations with stellar mass, initial mass function, metallicity, binary separation \citep{2017Moe,2017Sana,2021Niu}. However, our result is not very sensitive to these assumptions.
We find the SN environmental stars can be well fitted with two model stellar populations with mean log-ages of log($\tau$)=7.07$^{+0.03}_{-0.04}$ and 7.81$^{+0.04}_{-0.05}$ (i.e. 12 and 65~Myr, respectively) and a host-galaxy extinction of $A_V^{\rm host} = 0.31$ mag.

Their color-magnitude diagram (CMD) are plotted in Figure \ref{fig:progenitor}. 
We note that the earlier star-forming epoch (log($\tau$/yr)=7.81) is too old to be consistent with a core-collapse SN. Therefore, the SN progenitor is most likely to arise from the younger stellar population.
Note that the partial stripping of the H envelope by binary interaction has a very small effect on the core structure and stellar age \citep{2021Laplace}. 
Therefore, the age allows us to estimate a final luminosity of the progenitor base on the \textsc{parsec} single-star models.
Combing with the result of SED fitting (Section \ref{sec31}), the most probable position in the HRD for the progenitor is demonstrated with a red shade, which corresponds to a yellow supergiant with log($T_{\rm eff}/K)=3.72\pm0.08$ and log($L/L_{\odot})=5.17\pm0.04$ and an initial mass of about 16 $M_{\odot}$.
Considering the large difference between the derived total extinction of the SN progenitor ($A_V\sim3.12$~mag) and SN~2017gkk itself (0.073~mag), we suggest that the progenitor suffered from a non-negligible circumstellar extinction of $A_V^{\rm CSM}\sim3.05$~mag and circumstellar dust was rapidly destroyed the explosion.

\section{Summary and Discussion}\label{sec:dis}

\begin{figure} 
    \centering 
    \includegraphics[width=\linewidth]{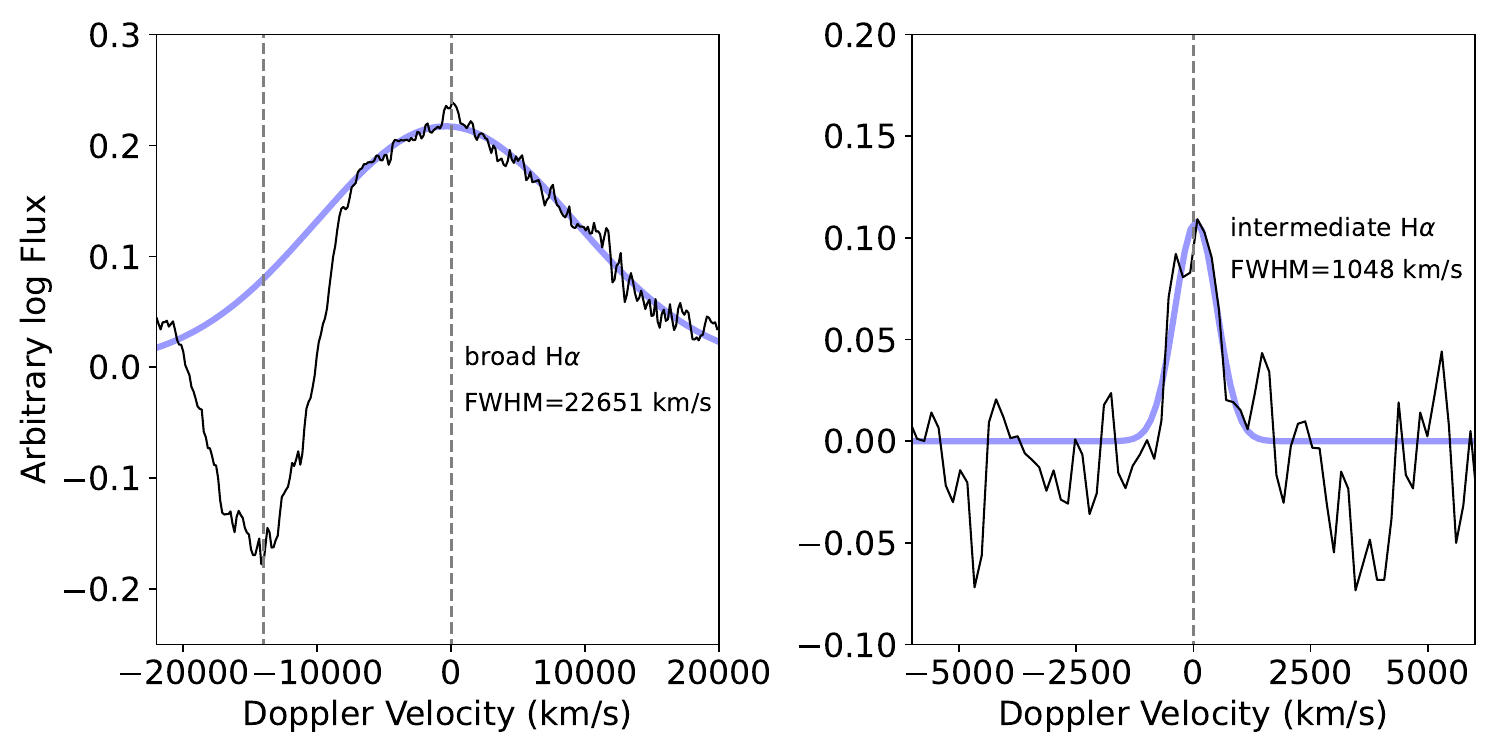}
    \caption{Signature for extended CSM from the H$\alpha$ profile of SN~2017gkk taken at about 17 days after explosion. In the left-hand panel, the H$\alpha$ profile is compared with a broad Gaussian function with FWHM$\sim$2.27$\times 10^4$~km~s$^{-1}$, and the blue-shifted absorption trough is around $-1.4 \times 10^4$~km~s$^{-1}$. The right-hand panel shows the H$\alpha$ residual after subtracting the broad Gaussian component, which can be fitted with an narrower Gaussian function of FWHM$\sim10^3$~km~s$^{-1}$.
    \label{fig:ha}}
\end{figure}

In this work, we report the discovery of a progenitor candidate of the Type~IIb SN~2017gkk in its pre-explosion HST images. We derived its properties by analyzing its SED and environment. We suggest that the progenitor may have significant circumstellar dust and the result of SED fitting is very sensitive to the adopted extinction.
With environmental analysis, however, we are able to constrain its age to be most likely 12~Myr. Combining both analysis, we suggest that the progenitor is most likely a YSG with $M_{\rm ini} \sim 15 M_{\odot}$.

The presence of CSM around SN~2017gkk is also indicated by a narrow emission line in its spectra. 
Figure \ref{fig:ha} shows the H$\alpha$ feature in its day-17 spectrum.
The velocity of the blue-shifted absorption trough is $-1.4 \times 10^4$~km~s$^{-1}$, and the red wing of the P Cygni profile matches a Gaussian shape with FWHM $\sim2.27 \times 10^4$~km~s$^{-1}$.
In addition to this broad component, which traces the fast expanding SN ejecta, there is also a narrower component with a FWHM of $\sim 10^3$~km~s$^{-1}$.
At the time this spectrum was taken, this component is most likely to arise from a blending zone, where the CSM has been swept up and accelerated by the forward shock; therefore the FWHM of the narrow component appears larger than the velocity of unshocked CSM \citep{2017hsnSmith}. 
Similar sign of CSM interaction was also observed for Type~IIb SN~2020bio \citep{2023Pellegrino}. The next spectrum of SN~2017gkk was observed 7 days later, when the narrow H$\alpha$ emission had almost faded.

Our discovery adds to the list of Type~IIb SNe with direct progenitor detections.
The initial mass of SN~2017gkk progenitor is very similar to those of the other 5 previously detected progenitors. Its effective temperature is close to that of SN~2011dh, and much higher than those of SNe~1993J and 2013df. 
It is worth mentioning that a single star of $M_{\rm ini} \sim 15 M_{\odot}$ will still retain a massive H envelope and will die as a Type~IIP, instead of Type~IIb, SN. 
Thus, our result supports an interacting binary progenitor channel for Type~IIb SNe (e.g. \citealp{2017Yoon}).

By the time of the last HST observation, SN~2017gkk was still brighter than the progenitor candidate. 
Future HST imaging is required to confirm whether this candidate is the genuine SN progenitor and has disappeared after explosion \citep{2014Maund,2023VanDyk}.
According to theoretical simulations, a companion star (most likely of OB-type) is expected to remain after the explosion of Type~IIb SN progenitor \citep[i.e.][]{2020Sravan}. The survived companions have been found for SN~1993J \citep{2004Maund}, SN~2001ig \citep{2018Ryder} and SN~2011dh \citep{2014Folatelli,2019Maund}.
All these companions are very hot stars with  log($T_{\rm eff}/K$)$\sim$ 4.3; but their initial masses can range from 9--14 $M_{\odot}$.
For SN~2017gkk, late imaging may also reveal its putative binary companion, which will provide important information for the progenitor's evolution.

\section*{acknowledgments}
This work is supported by the Strategic Priority Research Program of the Chinese Academy of Sciences, Grant No. XDB0550300. ZXN acknowledges support from the NSFC through grant No. 12303039, and NCS’s research is funded by the NSFC grants No. 12303051 and No. 12261141690. JFL acknowledges support from the NSFC through grants No. 11988101 and No. 11933004 and from the New Cornerstone Science Foundation through the New Cornerstone Investigator Program and the XPLORER PRIZE. 
This research is based on observations made with the NASA/ESA Hubble Space Telescope obtained from the Space Telescope Science Institute, which is operated by the Association of Universities for Research in Astronomy, Inc., under NASA contract NAS 5–26555. These observations are associated with programs 6359, 9042, 14668, 15166, and 16179. 
The specific observations analyzed can be accessed via \dataset[10.17909/qyxa-mn84]{https://doi.org/10.17909/qyxa-mn84}.

\vspace{5mm}

\bibliography{sample631}{}

\begin{thebibliography}{}
\expandafter\ifx\csname natexlab\endcsname\relax\def\natexlab#1{#1}\fi
\providecommand{\url}[1]{\href{#1}{#1}}
\providecommand{\dodoi}[1]{doi:~\href{http://doi.org/#1}{\nolinkurl{#1}}}
\providecommand{\doeprint}[1]{\href{http://ascl.net/#1}{\nolinkurl{http://ascl.net/#1}}}
\providecommand{\doarXiv}[1]{\href{https://arxiv.org/abs/#1}{\nolinkurl{https://arxiv.org/abs/#1}}}

\bibitem[{{Aldering} {et~al.}(1994){Aldering}, {Humphreys}, \& {Richmond}}]{1994Aldering}
{Aldering}, G., {Humphreys}, R.~M., \& {Richmond}, M. 1994, \aj, 107, 662, \dodoi{10.1086/116886}

\bibitem[{{Benvenuto} {et~al.}(2013){Benvenuto}, {Bersten}, \& {Nomoto}}]{2013Benvenuto}
{Benvenuto}, O.~G., {Bersten}, M.~C., \& {Nomoto}, K. 2013, \apj, 762, 74, \dodoi{10.1088/0004-637X/762/2/74}

\bibitem[{{Bersten} {et~al.}(2012){Bersten}, {Benvenuto}, {Nomoto}, {Ergon}, {Folatelli}, {Sollerman}, {Benetti}, {Botticella}, {Fraser}, {Kotak}, {Maeda}, {Ochner}, \& {Tomasella}}]{2012Bersten}
{Bersten}, M.~C., {Benvenuto}, O.~G., {Nomoto}, K., {et~al.} 2012, \apj, 757, 31, \dodoi{10.1088/0004-637X/757/1/31}

\bibitem[{{Bressan} {et~al.}(2012){Bressan}, {Marigo}, {Girardi}, {Salasnich}, {Dal Cero}, {Rubele}, \& {Nanni}}]{parsec.ref}
{Bressan}, A., {Marigo}, P., {Girardi}, L., {et~al.} 2012, \mnras, 427, 127, \dodoi{10.1111/j.1365-2966.2012.21948.x}

\bibitem[{{Cardelli} {et~al.}(1989){Cardelli}, {Clayton}, \& {Mathis}}]{ebvlaw}
{Cardelli}, J.~A., {Clayton}, G.~C., \& {Mathis}, J.~S. 1989, \apj, 345, 245, \dodoi{10.1086/167900}

\bibitem[{{Castelli} \& {Kurucz}(2003)}]{ck04}
{Castelli}, F., \& {Kurucz}, R.~L. 2003, in Modelling of Stellar Atmospheres, ed. N.~{Piskunov}, W.~W. {Weiss}, \& D.~F. {Gray}, Vol. 210, A20, \dodoi{10.48550/arXiv.astro-ph/0405087}

\bibitem[{{Claeys} {et~al.}(2011){Claeys}, {de Mink}, {Pols}, {Eldridge}, \& {Baes}}]{2011Claeys}
{Claeys}, J.~S.~W., {de Mink}, S.~E., {Pols}, O.~R., {Eldridge}, J.~J., \& {Baes}, M. 2011, \aap, 528, A131, \dodoi{10.1051/0004-6361/201015410}

\bibitem[{{Crockett} {et~al.}(2008){Crockett}, {Eldridge}, {Smartt}, {Pastorello}, {Gal-Yam}, {Fox}, {Leonard}, {Kasliwal}, {Mattila}, {Maund}, {Stephens}, \& {Danziger}}]{2008Crockett}
{Crockett}, R.~M., {Eldridge}, J.~J., {Smartt}, S.~J., {et~al.} 2008, \mnras, 391, L5, \dodoi{10.1111/j.1745-3933.2008.00540.x}

\bibitem[{{Dolphin}(2000)}]{2000Dolphin}
{Dolphin}, A.~E. 2000, \pasp, 112, 1383, \dodoi{10.1086/316630}

\bibitem[{{Farrell} {et~al.}(2020){Farrell}, {Groh}, {Meynet}, \& {Eldridge}}]{2020Farrell}
{Farrell}, E.~J., {Groh}, J.~H., {Meynet}, G., \& {Eldridge}, J.~J. 2020, \mnras, 494, L53, \dodoi{10.1093/mnrasl/slaa035}

\bibitem[{{Filippenko}(1988)}]{1988Filippenko}
{Filippenko}, A.~V. 1988, \aj, 96, 1941, \dodoi{10.1086/114940}

\bibitem[{{Folatelli} {et~al.}(2015){Folatelli}, {Bersten}, {Kuncarayakti}, {Benvenuto}, {Maeda}, \& {Nomoto}}]{2015Folatelli}
{Folatelli}, G., {Bersten}, M.~C., {Kuncarayakti}, H., {et~al.} 2015, \apj, 811, 147, \dodoi{10.1088/0004-637X/811/2/147}

\bibitem[{{Folatelli} {et~al.}(2014){Folatelli}, {Bersten}, {Benvenuto}, {Van Dyk}, {Kuncarayakti}, {Maeda}, {Nozawa}, {Nomoto}, {Hamuy}, \& {Quimby}}]{2014Folatelli}
{Folatelli}, G., {Bersten}, M.~C., {Benvenuto}, O.~G., {et~al.} 2014, \apjl, 793, L22, \dodoi{10.1088/2041-8205/793/2/L22}

\bibitem[{{Foreman-Mackey} {et~al.}(2013){Foreman-Mackey}, {Hogg}, {Lang}, \& {Goodman}}]{2013emcee}
{Foreman-Mackey}, D., {Hogg}, D.~W., {Lang}, D., \& {Goodman}, J. 2013, \pasp, 125, 306, \dodoi{10.1086/670067}

\bibitem[{{Itagaki}(2017)}]{2017Itagaki}
{Itagaki}, K. 2017, Transient Name Server Discovery Report, 2017-940, 1

\bibitem[{{Jerkstrand} {et~al.}(2015){Jerkstrand}, {Ergon}, {Smartt}, {Fransson}, {Sollerman}, {Taubenberger}, {Bersten}, \& {Spyromilio}}]{2015Jerkstrand}
{Jerkstrand}, A., {Ergon}, M., {Smartt}, S.~J., {et~al.} 2015, \aap, 573, A12, \dodoi{10.1051/0004-6361/201423983}

\bibitem[{{Kilpatrick} {et~al.}(2022){Kilpatrick}, {Coulter}, {Foley}, {Piro}, {Rest}, {Rojas-Bravo}, \& {Siebert}}]{2022Kilpatrick}
{Kilpatrick}, C.~D., {Coulter}, D.~A., {Foley}, R.~J., {et~al.} 2022, \apj, 936, 111, \dodoi{10.3847/1538-4357/ac8a4c}

\bibitem[{Koivisto {et~al.}(2022)Koivisto, Mattila, Kuncarayakti, \& Nagao}]{koivisto2022observational}
Koivisto, B. S.~N., Mattila, S., Kuncarayakti, D.~H., \& Nagao, T. 2022, Astronomy

\bibitem[{{Kourkchi} {et~al.}(2020){Kourkchi}, {Tully}, {Eftekharzadeh}, {Llop}, {Courtois}, {Guinet}, {Dupuy}, {Neill}, {Seibert}, {Andrews}, {Chuang}, {Danesh}, {Gonzalez}, {Holthaus}, {Mokelke}, {Schoen}, \& {Urasaki}}]{2020Kourkchi}
{Kourkchi}, E., {Tully}, R.~B., {Eftekharzadeh}, S., {et~al.} 2020, \apj, 902, 145, \dodoi{10.3847/1538-4357/abb66b}

\bibitem[{{Laplace} {et~al.}(2021){Laplace}, {Justham}, {Renzo}, {G{\"o}tberg}, {Farmer}, {Vartanyan}, \& {de Mink}}]{2021Laplace}
{Laplace}, E., {Justham}, S., {Renzo}, M., {et~al.} 2021, \aap, 656, A58, \dodoi{10.1051/0004-6361/202140506}

\bibitem[{{Li} {et~al.}(2024){Li}, {Hu}, {Li}, {Yang}, {Wang}, {Yan}, {Hu}, {Zhang}, {Mao}, {Riise}, {Gao}, {Sun}, {Liu}, {Xiong}, {Wang}, {Mo}, {Iskandar}, {Xi}, {Xiang}, {Wang}, {Sun}, {Zhang}, {Chen}, {Lin}, {Guo}, {Liu}, {Cai}, {Zhou}, {Zhao}, {Chen}, {Zheng}, {Li}, {Zhang}, {Xu}, {Lyu}, {Castro-Tirado}, {Chufarin}, {Potapov}, {Ionov}, {Korotkiy}, {Nazarov}, {Sokolovsky}, {Hamann}, \& {Herman}}]{2024Li}
{Li}, G., {Hu}, M., {Li}, W., {et~al.} 2024, \nat, 627, 754, \dodoi{10.1038/s41586-023-06843-6}

\bibitem[{{Maund}(2019)}]{2019Maund}
{Maund}, J.~R. 2019, \apj, 883, 86, \dodoi{10.3847/1538-4357/ab2386}

\bibitem[{{Maund} \& {Ramirez-Ruiz}(2016)}]{Maund2016}
{Maund}, J.~R., \& {Ramirez-Ruiz}, E. 2016, \mnras, 456, 3175, \dodoi{10.1093/mnras/stv2760}

\bibitem[{{Maund} {et~al.}(2014){Maund}, {Reilly}, \& {Mattila}}]{2014Maund}
{Maund}, J.~R., {Reilly}, E., \& {Mattila}, S. 2014, \mnras, 438, 938, \dodoi{10.1093/mnras/stt2131}

\bibitem[{{Maund} {et~al.}(2004){Maund}, {Smartt}, {Kudritzki}, {Podsiadlowski}, \& {Gilmore}}]{2004Maund}
{Maund}, J.~R., {Smartt}, S.~J., {Kudritzki}, R.~P., {Podsiadlowski}, P., \& {Gilmore}, G.~F. 2004, \nat, 427, 129, \dodoi{10.1038/nature02161}

\bibitem[{{Maund} {et~al.}(2011){Maund}, {Fraser}, {Ergon}, {Pastorello}, {Smartt}, {Sollerman}, {Benetti}, {Botticella}, {Bufano}, {Danziger}, {Kotak}, {Magill}, {Stephens}, \& {Valenti}}]{2011Maund}
{Maund}, J.~R., {Fraser}, M., {Ergon}, M., {et~al.} 2011, \apjl, 739, L37, \dodoi{10.1088/2041-8205/739/2/L37}

\bibitem[{{Moe} \& {Di Stefano}(2017)}]{2017Moe}
{Moe}, M., \& {Di Stefano}, R. 2017, \apjs, 230, 15, \dodoi{10.3847/1538-4365/aa6fb6}

\bibitem[{{Niu} {et~al.}(2021){Niu}, {Yuan}, {Wang}, \& {Liu}}]{2021Niu}
{Niu}, Z., {Yuan}, H., {Wang}, S., \& {Liu}, J. 2021, \apj, 922, 211, \dodoi{10.3847/1538-4357/ac2573}

\bibitem[{{Nomoto} {et~al.}(1995){Nomoto}, {Iwamoto}, \& {Suzuki}}]{1995Nomoto}
{Nomoto}, K.~I., {Iwamoto}, K., \& {Suzuki}, T. 1995, \physrep, 256, 173, \dodoi{10.1016/0370-1573(94)00107-E}

\bibitem[{{Onori}(2017)}]{2017Onori}
{Onori}, F. 2017, Transient Name Server Classification Report, 2017-964, 1

\bibitem[{{Pellegrino} {et~al.}(2023){Pellegrino}, {Hiramatsu}, {Arcavi}, {Howell}, {Bostroem}, {Brown}, {Burke}, {Elias-Rosa}, {Itagaki}, {Kaneda}, {McCully}, {Modjaz}, {Padilla Gonzalez}, {Pritchard}, \& {Yesmin}}]{2023Pellegrino}
{Pellegrino}, C., {Hiramatsu}, D., {Arcavi}, I., {et~al.} 2023, \apj, 954, 35, \dodoi{10.3847/1538-4357/ace595}

\bibitem[{{Podsiadlowski} {et~al.}(1993){Podsiadlowski}, {Hsu}, {Joss}, \& {Ross}}]{1993Podsiadlowski}
{Podsiadlowski}, P., {Hsu}, J.~J.~L., {Joss}, P.~C., \& {Ross}, R.~R. 1993, \nat, 364, 509, \dodoi{10.1038/364509a0}

\bibitem[{{Riess} {et~al.}(2022){Riess}, {Yuan}, {Macri}, {Scolnic}, {Brout}, {Casertano}, {Jones}, {Murakami}, {Anand}, {Breuval}, {Brink}, {Filippenko}, {Hoffmann}, {Jha}, {D'arcy Kenworthy}, {Mackenty}, {Stahl}, \& {Zheng}}]{H0.ref}
{Riess}, A.~G., {Yuan}, W., {Macri}, L.~M., {et~al.} 2022, \apjl, 934, L7, \dodoi{10.3847/2041-8213/ac5c5b}

\bibitem[{{Ryder} {et~al.}(2018){Ryder}, {Van Dyk}, {Fox}, {Zapartas}, {de Mink}, {Smith}, {Brunsden}, {Azalee Bostroem}, {Filippenko}, {Shivvers}, \& {Zheng}}]{2018Ryder}
{Ryder}, S.~D., {Van Dyk}, S.~D., {Fox}, O.~D., {et~al.} 2018, \apj, 856, 83, \dodoi{10.3847/1538-4357/aaaf1e}

\bibitem[{{Salpeter}(1955)}]{imf.ref}
{Salpeter}, E.~E. 1955, \apj, 121, 161, \dodoi{10.1086/145971}

\bibitem[{{Sana}(2017)}]{2017Sana}
{Sana}, H. 2017, in The Lives and Death-Throes of Massive Stars, ed. J.~J. {Eldridge}, J.~C. {Bray}, L.~A.~S. {McClelland}, \& L.~{Xiao}, Vol. 329, 110--117, \dodoi{10.1017/S1743921317003209}

\bibitem[{{Schlafly} \& {Finkbeiner}(2011)}]{2011sfd}
{Schlafly}, E.~F., \& {Finkbeiner}, D.~P. 2011, \apj, 737, 103, \dodoi{10.1088/0004-637X/737/2/103}

\bibitem[{{Smith}(2017)}]{2017hsnSmith}
{Smith}, N. 2017, in Handbook of Supernovae, ed. A.~W. {Alsabti} \& P.~{Murdin}, 403, \dodoi{10.1007/978-3-319-21846-5_38}

\bibitem[{{Sorce} {et~al.}(2014){Sorce}, {Tully}, {Courtois}, {Jarrett}, {Neill}, \& {Shaya}}]{2014Sorce}
{Sorce}, J.~G., {Tully}, R.~B., {Courtois}, H.~M., {et~al.} 2014, \mnras, 444, 527, \dodoi{10.1093/mnras/stu1450}

\bibitem[{{Sravan} {et~al.}(2020){Sravan}, {Marchant}, {Kalogera}, {Milisavljevic}, \& {Margutti}}]{2020Sravan}
{Sravan}, N., {Marchant}, P., {Kalogera}, V., {Milisavljevic}, D., \& {Margutti}, R. 2020, \apj, 903, 70, \dodoi{10.3847/1538-4357/abb8d5}

\bibitem[{{STScI Development Team}(2013)}]{pysynphot.ref}
{STScI Development Team}. 2013, {pysynphot: Synthetic photometry software package}, Astrophysics Source Code Library, record ascl:1303.023.
\newblock \doeprint{1303.023}

\bibitem[{{Sun} {et~al.}(2021){Sun}, {Maund}, {Crowther}, {Fang}, \& {Zapartas}}]{Sun2021}
{Sun}, N.-C., {Maund}, J.~R., {Crowther}, P.~A., {Fang}, X., \& {Zapartas}, E. 2021, \mnras, 504, 2253, \dodoi{10.1093/mnras/stab994}

\bibitem[{{Tully} {et~al.}(2016){Tully}, {Courtois}, \& {Sorce}}]{2016Tully}
{Tully}, R.~B., {Courtois}, H.~M., \& {Sorce}, J.~G. 2016, \aj, 152, 50, \dodoi{10.3847/0004-6256/152/2/50}

\bibitem[{{Tully} {et~al.}(2013){Tully}, {Courtois}, {Dolphin}, {Fisher}, {H{\'e}raudeau}, {Jacobs}, {Karachentsev}, {Makarov}, {Makarova}, {Mitronova}, {Rizzi}, {Shaya}, {Sorce}, \& {Wu}}]{2013Tully}
{Tully}, R.~B., {Courtois}, H.~M., {Dolphin}, A.~E., {et~al.} 2013, \aj, 146, 86, \dodoi{10.1088/0004-6256/146/4/86}

\bibitem[{{Van Dyk} {et~al.}(2014){Van Dyk}, {Zheng}, {Fox}, {Cenko}, {Clubb}, {Filippenko}, {Foley}, {Miller}, {Smith}, {Kelly}, {Lee}, {Ben-Ami}, \& {Gal-Yam}}]{2014VanDyk}
{Van Dyk}, S.~D., {Zheng}, W., {Fox}, O.~D., {et~al.} 2014, \aj, 147, 37, \dodoi{10.1088/0004-6256/147/2/37}

\bibitem[{{Van Dyk} {et~al.}(2023){Van Dyk}, {de Graw}, {Baer-Way}, {Zheng}, {Filippenko}, {Fox}, {Smith}, {Brink}, {de Jaeger}, {Kelly}, \& {Vasylyev}}]{2023VanDyk}
{Van Dyk}, S.~D., {de Graw}, A., {Baer-Way}, R., {et~al.} 2023, \mnras, 519, 471, \dodoi{10.1093/mnras/stac3549}

\bibitem[{{Weisz} {et~al.}(2024){Weisz}, {Dolphin}, {Savino}, {McQuinn}, {Newman}, {Williams}, {Kallivayalil}, {Anderson}, {Boyer}, {Correnti}, {Geha}, {Sandstrom}, {Cole}, {Warfield}, {Skillman}, {Cohen}, {Beaton}, {Bressan}, {Bolatto}, {Boylan-Kolchin}, {Brooks}, {Bullock}, {Conroy}, {Cooper}, {Dalcanton}, {Dotter}, {Fritz}, {Garling}, {Gennaro}, {Gilbert}, {Girardi}, {Johnson}, {Johnson}, {Kalirai}, {Kirby}, {Lang}, {Marigo}, {Richstein}, {Schlafly}, {Tollerud}, \& {Wetzel}}]{2024Weisz}
{Weisz}, D.~R., {Dolphin}, A.~E., {Savino}, A., {et~al.} 2024, \apjs, 271, 47, \dodoi{10.3847/1538-4365/ad2600}

\bibitem[{{Yoon} {et~al.}(2017){Yoon}, {Dessart}, \& {Clocchiatti}}]{2017Yoon}
{Yoon}, S.-C., {Dessart}, L., \& {Clocchiatti}, A. 2017, \apj, 840, 10, \dodoi{10.3847/1538-4357/aa6afe}

\end{thebibliography}
\bibliographystyle{aasjournal}

\end{document}